\documentclass[aps,prb,twocolumn,groupedaddress]{revtex4}
\usepackage{graphicx}
\usepackage{wasysym}
\usepackage{bm}
\usepackage{color}

\begin{document}
\title{ Magnetothermoelectric transport properties in phosphorene}
\author{R. Ma$^{1,2}$}
\email{njrma@hotmail.com}
\author{S. W. Liu$^{1}$}
\author{M. X. Deng$^{2}$}
\author{L. Sheng$^{2,3}$}
\email{shengli@nju.edu.cn}
\author{D. Y. Xing$^{2,3}$}
\email{dyxing@nju.edu.cn}
\author{D. N. Sheng$^{4}$}
\address{$^1$ Jiangsu Key Laboratory for Optoelectronic Detection of Atmosphere and Ocean,
  Nanjing University of Information Science and Technology, Nanjing 210044, China\\
$^2$ National Laboratory of Solid State Microstructures and
Department of Physics, Nanjing University, Nanjing 210093, China\\
$^3$ Collaborative Innovation Center of Advanced Microstructures,
Nanjing 210093, China\\
$^4$ Department of Physics and Astronomy, California State
University, Northridge, California 91330, USA}

\begin{abstract}
We numerically study the electrical and thermoelectric transport properties
in phosphorene in the presence of both a magnetic field and disorder.
The quantized Hall conductivity is similar to that of a conventional
two-dimensional electron gas, but the positions of all the Hall plateaus
shift to the left due to the spectral asymmetry, in agreement with
the experimental observations.
The thermoelectric conductivity and Nernst signal
exhibit remarkable anisotropy, and the thermopower is nearly isotropic.
When a bias voltage is applied between top and bottom layers of phosphorene,
both thermopower and Nernst signal are enhanced and their peak values become large.
\end{abstract}

%\pacs{73.50.-h; 72.10.-d; 73.50.Lw, 73.43.Cd}
\maketitle

\section{Introduction}
\label{sec:intro}

Recently, a new two-dimensional (2D) semiconductor material,
called black phosphorus, has attracted much attention because of
its unique electronic properties and potential applications
~\cite{Reich2014,Li2015,Li2014,Liu2014,Gomez2014,Xia2014,Koenig2014,Zare2017}.
Black phosphorus is a layered material, in which individual atomic
layers are stacked together by van der Waals interactions.
Similar to graphene, black phosphorus can be mechanically exfoliated to
obtain samples with a few or single layers, with the latter being known
as phosphorene~\cite{Li2014,Liu2014}.
Within a phosphorene sheet, every phosphorous atom is covalently
bonded with three neighboring atoms, forming a puckered honeycomb structure.
This hinge-like puckered structure
leads to a highly anisotropic electronic structure, with a direct band gap of
1.51 $eV$ that can be potentially tuned by changing the number of layers~\cite{Qiao2014}.
The low-energy dispersion is quadratic with very different effective masses
along armchair and zigzag directions for both electrons and holes~\cite{Fei2014}.
Under a strong perpendicular magnetic field, an integer quantum Hall effect (QHE)
has been realized in black phosphorus~\cite{Li2015}.
Field effect transistors based on a few layers of phosphorene are found to have a
higher on-off current ratio at room temperatures~\cite{Li2014,Liu2014,Gomez2014,Xia2014},
making it a promising candidate material for fabrication of switching devices.
On the other hand, the experimental measurements of the thermoelectric power
in bulk black phosphorus indicate that~\cite{Flores2015} the Seebeck coefficient is
335$\pm 10\mu V/K$ at room temperature, and it increases with temperature,
suggesting that phosphorene-based materials could be a good candidate
for thermoelectric applications.

Up to now, some theoretical investigations have been carried out on
the thermoelectric properties of phosphorene~\cite{Fei2014,Qin2014,Lv2014,Satoru2014}.
It is found that the thermoelectric performance of bulk black phosphorus
can be greatly enhanced by strain effect, and the thermopower exhibits
an anisotropic property at high temperatures~\cite{Fei2014,Qin2014}.
It has been pointed out that the Seebeck coefficient of phosphorene
is larger than that of bulk black phosphorus~\cite{Lv2014}.
So far, the effects of a strong magnetic field and disorder in phosphorene have not been
investigated. It is well known that when the magnetic field is absent,
the thermoelectric transport depends crucially on impurity scattering as well as
thermal activation. In a strong magnetic field, due to the fact that
high-degenerated Landau levels (LLs) dominate transport processes,
the thermoelectric properties in this unique anisotropic system may exhibit
complex physical properties.
On the other hand, in the experiment, Chang-Ran Wang $et$ $al.$~\cite{Wang2011}
demonstrate that the thermopower can be enhanced greatly at a low temperature
by using a dual-gated bilayer graphene device, which was predicted theoretically
as an effect of opening of a band gap~\cite{hao2010}.
Up to now, there have been no experimental studies on tuning the thermopower
of phosphorene. It is highly desirable to investigate disorder effect and
thermal activation on the thermoelectric transport for different transport
directions of phosphorene in the presence of a strong perpendicular magnetic field.

In this paper, we carry out a numerical study on
the electrical and thermoelectric transport
of phosphorene in the presence of a strong magnetic field and disorder.
We investigate the effects of disorder and thermal activation on the
broadening of LLs and the corresponding thermoelectric
transport coefficients.
We show that the quantized Hall conductivity of phosphorene is similar to
that of a conventional two-dimensional electron gas (2DEG),
but the positions of all the Hall plateaus shift to the left due to
the spectral asymmetry, in agreement with the experimental observations.
Interestingly, both the thermoelectric conductivities and Nernst signal
exhibit remarkable anisotropy, but thermopower is nearly isotropic.
When a bias voltage is applied between the top and bottom layers of phosphorene,
it is interesting to find that both the thermopower and
Nernst signal are enhanced compared to the unbiased case.
These features can be understood as being due to the increase of the bulk energy gap.
Moreover, we also study the disorder effect
on the electrical and thermoelectric transport in phosphorene.
With increasing disorder strength, the Hall plateaus can be destroyed
through the float-up of extended levels toward the band center and
higher plateaus disappear first.
The $\nu=0$ Hall plateau is most robust against disorder scattering.
In the presence of the strong magnetic field,
both thermopower and Nernst signal are robust to the disorder,
because of the existence of the quantized LLs.

This paper is organized as follows. In Sec.\ II, the model Hamiltonian
of phosphorene is introduced. In Sec.\ III, numerical results of the electrical
and thermoelectric transport coefficients obtained by using exact diagonalization
are presented. The final section contains a summary.

\section{Model and Methods}

\begin{figure}[tbh]
\par
\includegraphics[width=2.9in]{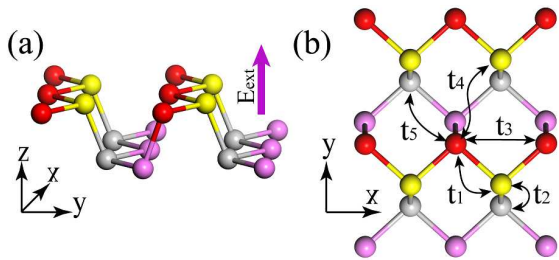}
\caption{(color online). (a) Crystal structure of phosphorene, and
(b) the top view of phosphorene, where the hopping parameters $t_{ij}$
of the tight-binding model are indicated.
Circles of different colors correspond to atoms located in different
planes within a single puckered layer. } \label{fig.1}
\end{figure}

Phosphorene has a characteristic puckered structure as shown
in Fig.\ref{fig.1}, which leads to the two anisotropic in-plane directions.
To specify the system size, we assume that the sample has totally $L_{y}$ zigzag
chains with $L_{x}$ atomic sites on each zigzag chain~\cite{Sheng2006}.
The total number of sites in the sample is denoted as
$N=L_{x}\times L_{y}$. In our numerical calculation,
the system size is taken to be $N=96\times 48$,
and the distance between nearest-neighbor sites is chosen
as the unit of length.
It have been shown that the calculated results do not depend on
the system sizes, as long as the system lengths are reasonably large~\cite{Ma2009}.
The unit cell of phosphorene contains four phosphorus atoms,
with two phosphorus atoms on the top layer and the other
two on the bottom layer.
When a magnetic field is applied
perpendicular to the phosphorene film, the Hamiltonian can
be written in the tight-binding form~\cite{Rudenko2014},
\begin{eqnarray}
H&=&\sum\limits_{\langle {ij}\rangle}t_{ij}e^{ia_{ij}}c_{i}^{\dagger
}c_{j}+\sum\limits_{i}
U_i c_{i}^{\dagger}c_{i}+H.c.
+\sum\limits_{i}w_i c_{i}^{\dagger}c_{i}\ .
\end{eqnarray}
Here, the summation of $\langle {ij}\rangle$
runs over neighboring lattice sites, and $c_{i}^{\dagger}$ and $c_{i}$
are the creation and annihilation operators of electrons
on site $i$. The hopping integrals $t_{ij}$ between
site $i$ and its neighbours $j$ are described in Fig.\ref{fig.1}.
The hopping integral $t_{1}$ corresponds to the connection
along a zigzag chain in the upper or lower layer,
and $t_{2}$ stands for the connection between a pair of
zigzag chains in the upper and lower layers. $t_3$
is between the nearest-neighbour sites of a pair of zigzag chains
in the upper or lower layer, and $t_4$ is between
the next nearest-neighbour sites of a pair of zigzag chains
in the upper and lower layers. $t_5$ is the hopping integral
between two atoms on upper and lower zigzag chains that are farthest
from each other. The values of these hopping integrals are
$t_1= -1.220$ $eV$, $t_2=3.665$ $eV$, $t_3=-0.205$ $eV$, $t_4= -0.105$ $eV$,
and $t_5= -0.055$ $eV$~\cite{Rudenko2014}.
The magnetic flux per hexagon $\phi =\sum_{{\small {\mbox{\hexagon}}}}a_{ij}=\frac{2\pi}{M}$
is proportional to the strength of the applied magnetic field $B$,
where $M$ is an integer and the lattice constant
is taken to be unity.
In the presence of a uniform perpendicular electric field,
the electrostatic potentials of
the top and bottom layers  are set as
$U_{top}=-U_{bottom}=\frac{1}{2}\Delta_g$~\cite{Ma2016}.
For illustrative purpose, a relatively large potential difference
$\Delta_g$=2$\vert t_1 \vert$ is taken.
The last term is the on-site random potential accounting for Anderson
disorder, where $w_i$ is assumed to be uniformly distributed in the range
$w_i\in \lbrack -W/2,W/2]$, with $W$ as the disorder strength~\cite{Huo1992,Sheng97}.

In the linear response regime, the charge current in response to an
electric field or a temperature gradient can be written as  ${\bf J}
= {\hat \sigma} {\bf E} + {\hat \alpha} (-\nabla T)$, where ${\hat
\sigma}$ and ${\hat \alpha}$ are the electrical and thermoelectric
conductivity tensors, respectively. The electrical conductivity $\sigma_{ji}$
at zero temperature can be calculated by using the Kubo formula
\begin{eqnarray}
%\[
\sigma _{ji}= \frac{ie^{2}\hbar}{A}\sum_{{\epsilon
_{\alpha}}\neq{\epsilon _{\beta}}}\frac{f(\epsilon_\alpha)-f(\epsilon_\beta)}{\epsilon_\alpha-\epsilon_\beta}\frac{\langle
\alpha\mid V_j\mid\beta\rangle\langle\beta\mid V_i\mid\alpha
\rangle}{\epsilon_\alpha-\epsilon_\beta+i\eta},
%\]
\end{eqnarray}
Here, $\epsilon_\alpha$ and $\epsilon_\beta$ are the eigenenergies
corresponding to the eigenstates $\vert\alpha\rangle$ and
$\vert\beta\rangle$ of the system, respectively, which can be obtained through exact
diagonalization of the Hamiltonian Eq.\ (1). $A$ is the
area of the sample, and $f(\epsilon_\alpha)$ and $f(\epsilon_\beta)$
are the Fermi-Dirac distribution functions, defined as
$f(x) = 1/[e^{(x-E_F)/k_B T}+1]$. $V_{j}$ and $V_{i}$ are
the velocity operators, and $\eta$ is the positive infinitesimal,
accounting for the finite broadening of the LLs.
With tuning Fermi energy $E_F$, a series of integer-quantized
Hall plateaus of $\sigma_{xy}$ appear, each one corresponding to $E_F$
moving in the gaps between two neighboring LLs.

\begin{figure}[tbh]
\includegraphics[width=2.9in]{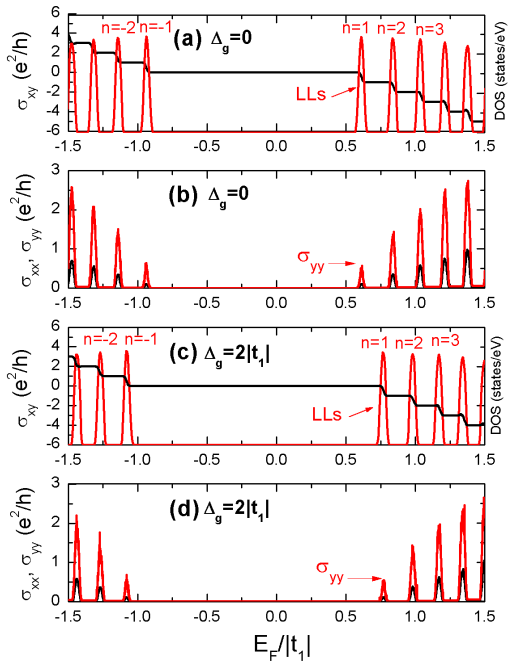}
\caption{ (color online). Calculated electron density of states,
Hall conductivity $\sigma_{xy}$
and longitudinal conductivity $\sigma_{xx}$ ($\sigma_{yy}$)
of phosphorene as functions of the Fermi energy at zero temperature.
(a)-(b) $\Delta_g=0$, (c)-(d) $\Delta_g=2\vert t_1 \vert$.
The system size is taken to be $N=96\times 48$, magnetic flux $\phi=2\pi/48$,
and disorder strength $W=0.5$. The positive infinitesimal $\eta$ is set to $10^{-3}eV$.
Each data point is obtained by averaging over up to $2000$ disorder configurations.
} \label{fig.2}
\end{figure}

\begin{figure*}[tbh]
\includegraphics[width=5.8in]{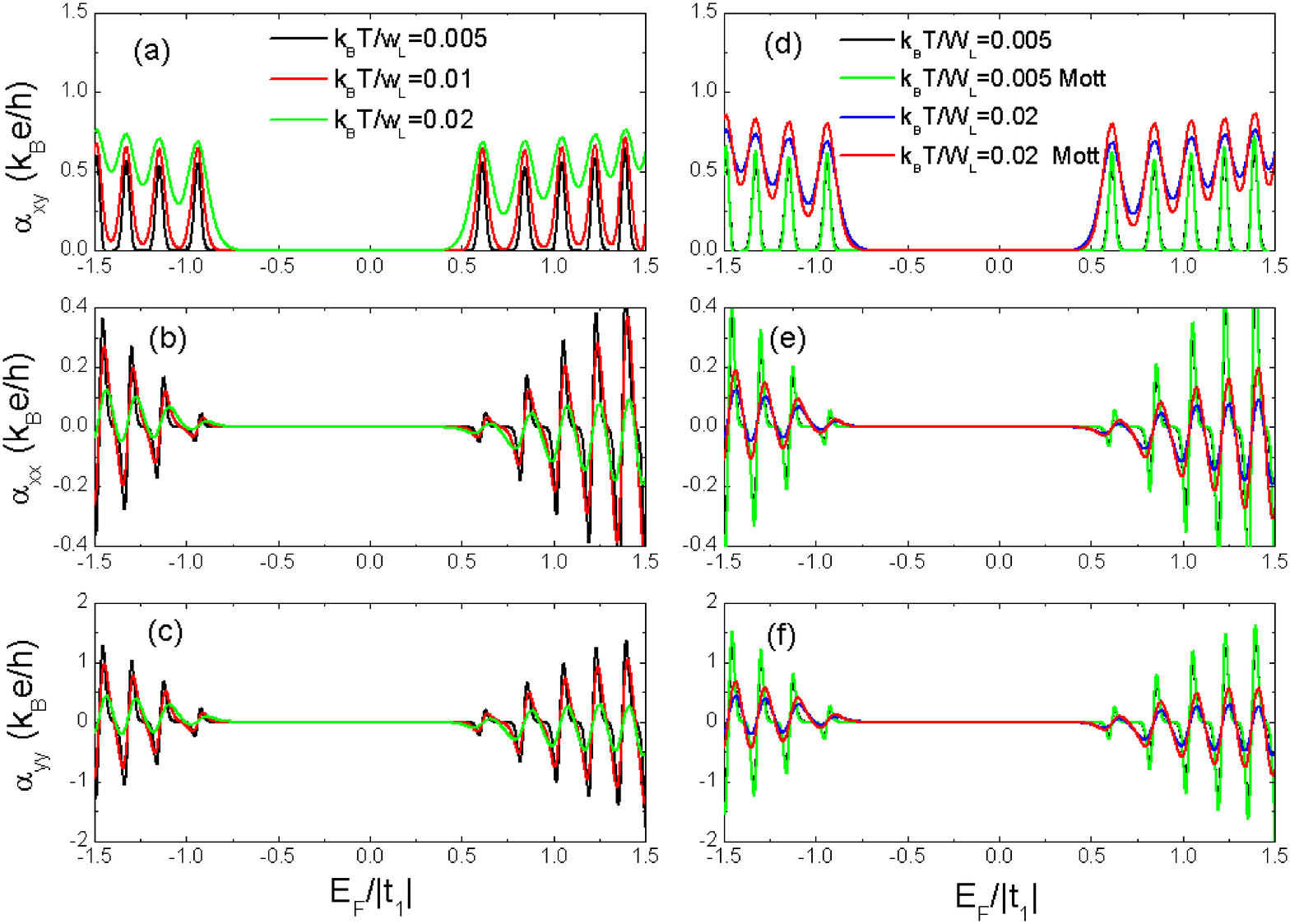}
\caption{ (color online). Thermoelectric conductivities at finite temperatures of phosphorene.
(a)-(c)$\alpha_{xy}(E_F, T)$, $\alpha_{xx}(E_F,T)$ and $\alpha_{yy}(E_F,T)$
as functions of the Fermi energy at different temperatures.
(d)-(f) Comparison of the results from numerical calculations and from
the generalized Mott relation for two characteristic temperatures,
$k_{B}T/W_L=0.005$ and $k_BT/W_L=0.02$.
Here, the asymmetric gap $W_L$ is equal to $W_L/\vert t_1 \vert=1.548$,
which is determined by the distance between two neighboring peaks of
$\sigma_{xx}$ peaks around zero energy.
The other parameters are chosen to be the same as in Fig.\ref{fig.2}.
} \label{fig.3}
\end{figure*}

We exactly diagonalize the model Hamiltonian in the presence of
disorder~\cite{Sheng97}, and obtain the transport coefficients
by using the energy spectra and wave functions. In practice,
we can first calculate the electrical conductivities $\sigma_{ji}$
at zero temperature, and then use the relation~\cite{Jonson84}
\begin{eqnarray}
\sigma_{ji}(E_F, T) &=& \int d\epsilon \,\sigma_{ji}(\epsilon)
\left ( - {\partial f(\epsilon) \over \partial \epsilon } \right), \\
\alpha_{ji}(E_F, T) &=& {-1\over eT} \int d\epsilon\,
\sigma_{ji}(\epsilon) (\epsilon-E_F) \left ( - {\partial f(\epsilon)
\over \partial \epsilon } \right), \label{eq:conductance-finiteT}
\end{eqnarray}
to obtain the electrical and thermoelectric conductivity at finite
temperatures. At low temperatures, the second equation can
be approximated as
\begin{equation}
\alpha_{ji}(E_F, T) =-\frac {\pi^2k_B^2T}{3e}\left. \frac
{d\sigma_{ji}(\epsilon, T)}{d\epsilon} \right|_{\epsilon =E_F},
\label{eq:Mott-relation}
\end{equation}
which is the semiclassical Mott relation~\cite{Jonson84,Oji84}. The
validity of this relation will be examined for the present phosphorene system.
The thermopower and Nernst signal can be
calculated subsequently from~\cite{footnote1}
\begin{eqnarray}
S_{xx} &=&  { E_x \over \nabla_x T} =({\sigma_{yy}\alpha_{xx}+\sigma_{xy}\alpha_{xy}})/{D}\\
S_{yy} &=&  { E_y \over \nabla_y T} =({\sigma_{xx}\alpha_{yy}+\sigma_{xy}\alpha_{xy}})/{D}\\
S_{xy} &=& { E_x \over \nabla_y T} =({\sigma_{yy}\alpha_{xy}-\sigma_{xy}\alpha_{yy}})/{D}\\
S_{yx} &=& { E_y \over \nabla_x T} =({-\sigma_{xx}\alpha_{xy}+\sigma_{xy}\alpha_{xx}})/{D},
\label{eq:thermoelectric}
\end{eqnarray}
with $D=\sigma_{xx}\sigma_{yy}+\sigma_{xy}^2$.

\begin{figure}[tbh]
\includegraphics[width=2.9in]{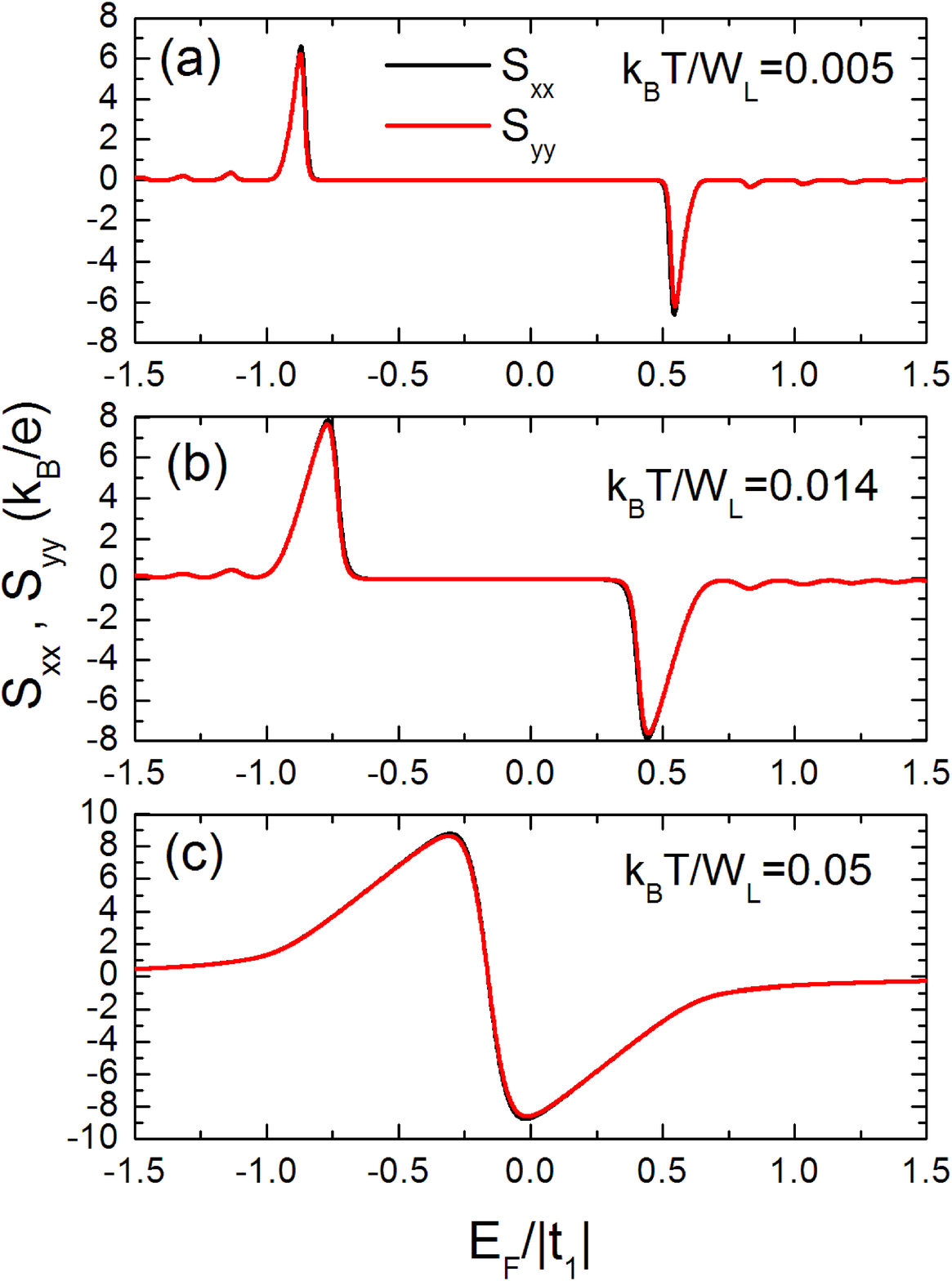}
\caption{ (color online). Calculated thermopower $S_{xx}$ and $S_{yy}$
as functions of the Fermi energy at three different temperatures in phosphorene.
} \label{fig.4}
\end{figure}

\begin{figure}[tbh]
\includegraphics[width=2.9in]{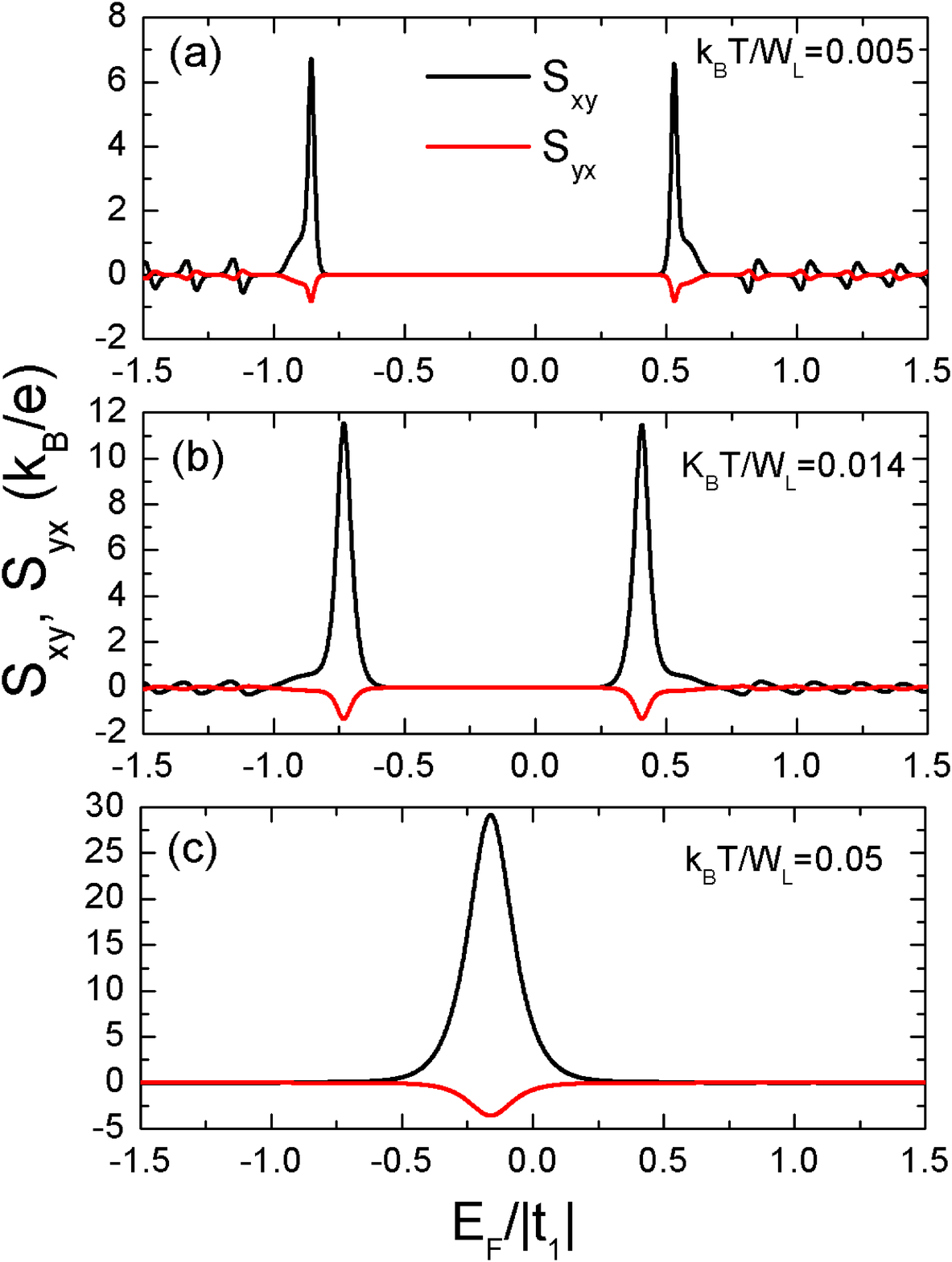}
\caption{ (color online). Calculated Nernst signal $S_{xy}$ and $S_{yx}$
as functions of the Fermi energy at three different temperatures in phosphorene.
} \label{fig.5}
\end{figure}

\section{Results and Discussion}

\subsection{The electrical and thermoelectric transport}

In Fig.\ref{fig.2}, we first show the electron density of states (DOS),
the Hall conductivity $\sigma_{xy}$ and longitudinal conductivity $\sigma_{xx}$
($\sigma_{yy}$) as functions of $E_F$ at zero temperature.
In the presence of a magnetic field, the DOS is discrete, forming a series of LLs,
as seen from the right part of Fig.\ref{fig.2}(a).
We will call the LL just above $E_F=0$ as $n=1$ LL,
that just below $E_F=0$ as $n=-1$ LL, and so on.
Clearly, the central $n=0$ LL around $E_F$=0 is markedly absent.
Moreover, all the LLs in the positive and negative regions are somewhat
asymmetric in position, which can be attributed to the absence of
particle-hole symmetry of the present band structure~\cite{Ma2016}.
The Hall conductivity is strictly quantized due to the quantized LLs.
As can be seen from Fig.\ref{fig.2}(a),
the Hall conductivity exhibits a sequence of plateaus at $\sigma _{xy}=\nu e^2/h$,
where the filling factor $\nu=\pm kg_{s}$ with $k$ as an integer, and $g_{s}=1$
due to the lack of the valley degeneracy.
With each additional LL being occupied, the total Hall conductivity
is increased by $e^{2}/h$. This is an invariant as long as the states
between the $n$-th and $(n-1)$-th LL are localized.
Around zero energy point, a pronounced plateau with $\nu=0$ is found,
which can only be understood as being due to the appearance of the bulk energy gap
between the valence and conduction bands~\cite{Rudenko2014}.
Moreover, one can also see that the width of the $\nu=0$ plateaus is determined
by the LL spacing between the two nearest levels.
These results are somewhat similar to the conventional integer
QHE found in the 2D semiconductor systems subject to a perpendicular magnetic field,
but the conductivity plateaus in the conduction band and valence band are not
antisymmetric in energy due to the asymmetric positions of the LLs.
Our calculated results are in good agreement with the experimental observation
of the QHE in black phosphorus~\cite{Li2015}.
In Figs.\ref{fig.2}(b), the longitudinal conductivity $\sigma_{xx}$ along the zigzag direction
shows some pronounced peaks when the Fermi energy coincides with the LLs.
According to the Kubo formula in Eq.\ (2), $\sigma_{xx}$ is proportional to
$\sigma_{xx}\propto n{\partial f(\epsilon) \over \partial \epsilon }\delta(\epsilon-\epsilon_n)$.
As a result, $\sigma_{xx}$ displays a very similar structure to that of the DOS.
In addition, the peak values of $\sigma_{xx}$ increase along with
the increase of $E_F$ because of the larger transmission rate
between the one-electron states $\vert\alpha\rangle$ and $\vert\beta\rangle$
with higher LLs index~\cite{Zhou2015,Ando1974,Vasilopoulos2012,Zhou2014}.
The results of the longitudinal conductivity along the armchair direction $\sigma_{yy}$
are qualitatively similar to those of $\sigma_{xx}$.
Interestingly, the longitudinal conductivity exhibits an obvious
anisotropic property, with the value along the zigzag direction
much smaller than that along the armchair direction $\sigma_{xx}<\sigma_{yy}$.
This anisotropic property is due to the anisotropic electronic structure~\cite{Lv2014,Fei2014}.
The band along the zigzag direction is much flatter than that along the armchair direction,
which results in the much larger band effective mass and therefore much
smaller carrier mobility and electrical conductivity in the
zigzag direction.

When a bias voltage or a potential difference $\Delta_g$ is applied between
the top and bottom layer of phosphorene, the Hall conductivity
exhibits some interesting features. In Figs.\ref{fig.2}(c)-(d), we show
the calculated DOS and electrical conductivity for a bias voltage
$\Delta_g=2\vert t_1 \vert$. As seen from Fig.\ref{fig.2}(c),
the gap between the $n=\pm 1$ LLs is increased,
which can only be understood as being due to the increase of the bulk energy gap
between the valence and conduction bands. The Hall conductivity
exhibits the same quantization rule as that of the unbiased phosphorene,
while the width of the $\nu=0$ Hall plateau is enlarged due to the increase
of the gap between the LLs. The longitudinal conductivity exhibits
similar anisotropic behavior to that of the unbiased case,
as shown in Figs.\ref{fig.2}(d).

Now we turn to study the thermoelectric transport coefficients of phosphorene.
In Fig.\ref{fig.3}, we first plot the calculated thermoelectric conductivity
at finite temperatures. Here, the temperature is defined by the ratio
between $k_BT$ and $W_L$, where $W_L$ is the energy difference between the two nearest
$\sigma_{xx}$ peaks around zero energy.
As shown in Figs.\ref{fig.3}(a)-(c),
the transverse thermoelectric conductivity $\alpha_{xy}$ displays
a series of peaks, while the longitudinal thermoelectric conductivity
${\alpha_{xx}}$ (${\alpha_{yy}}$) oscillates and changes sign
at the center of each LL.
As seen from Fig.\ref{fig.3}(a), $\alpha_{xy}$ displays a pronounced
valley with $\alpha_{xy}=0$ around zero energy at low temperatures.
These are consistent with the presence of $\nu=0$ Hall plateau due to
the lack of the valley degeneracy in phosphorene.
Moreover, the positions of all the peaks in $\alpha_{xy}$ are asymmetric
in energy due to the spectral asymmetry~\cite{Ma2016}.
In Figs.\ref{fig.3}(b) and (c),
the longitudinal thermoelectric conductivity $\alpha_{xx}$($\alpha_{yy}$)
also exhibits an obvious anisotropy,
with the value along the zigzag direction much smaller than that along
the armchair direction $\alpha_{xx}<\alpha_{yy}$. This is due to
the anisotropic longitudinal conductivity.
In Figs.\ref{fig.3}(d)-(f), we also compare the above results with those calculated
from the semiclassical Mott relation given in Eq.(\ref{eq:Mott-relation}).
The Mott relation is found to remain valid only at low temperatures,
indicating that the semiclassical Mott relation is asymptotically valid
in Landau-quantized systems, as suggested in Ref.~\onlinecite{Jonson84}.
With increasing temperature, its deviation will become more and more pronounced.
However, if we take into account the finite-temperature values of
electrical conductivity, the Mott relation still predicts the correct asymptotic behavior.

We further discuss some interesting features of the thermopower
and Nernst signal in phosphorene using Eqs.(6)-(9), which can be directly
observed in experiments by measuring the responsive electric fields.
In Fig. \ref{fig.4},
we first show the thermopower along the zigzag direction, $S_{xx}$,
at different temperatures.
As seen from Fig. \ref{fig.4}(a),
$S_{xx}$ exhibits a series of peaks at all the LLs.
The largest peak values of $S_{xx}$ at $n=\pm 1$ LLs are found to be
$\pm 6.6$ $k_B/e$ ($\pm 568.7$ $\mu V/K$) at $k_BT=0.005W_L$.
With the increase of temperature,
the peaks gradually rise and widen, their positions shifting towards
$E_F=-0.16\vert t_1 \vert$.
As seen from Fig. \ref{fig.4}(c),
the peak values around $E_F=-0.16\vert t_1 \vert$ increase to
$\pm 8.8$ $k_B/e$ ($\pm 758.3$ $\mu V/K$) at $k_BT=0.05W_L$.
We can see that there is only a very small difference between $S_{xx}$ and $S_{yy}$,
and so the thermopower is nearly isotropic,
which is qualitatively consistent with that obtained by Fei $et$ $al.$~\cite{Fei2014,Qin2014,Lv2014}.

In Fig. \ref{fig.5}, we show the Nernst signals $S_{xy}$ and $S_{yx}$
at different temperatures.
%The largest peak values of $S_{xy}$ at $n=\pm 1$ LLs is found to be
%6.7 $k_B/e$ (577.3 $\mu V/K$) at $k_BT=0.005W_L$.
With the increase of temperature,
the largest peak values of $S_{xy}$ and $S_{yx}$ increase, and
the peak positions shift towards $E_F=-0.16\vert t_1 \vert$.
%which is similar to those of the thermopower.
%However, unlike the isotropic features of the thermopower,
Interestingly, the Nernst signals exhibit
remarkable anisotropic property over the whole temperature range.
In addition to the opposite signs between $S_{xy}$ and $S_{yx}$,
there is a big difference in magnitude between them.
For example, at $k_BT=0.05W_L$,
the peak value of $S_{xy}$ is $29.2$ $k_B/e$ (2516.2 $\mu V/K$),
but the value of $S_{yx}$ is only $-3.6$ $k_B/e$ ($-310.2$ $\mu V/K$),
as seen from Fig.\ref{fig.5}(c).
This anisotropic origin can be understood by the following argument.
Our calculation shows that in Eq. (8) or (9), the first term in the bracket
is much greater than the second term, i.e.,
$\sigma_{yy}\alpha_{xy}\gg\sigma_{xy}\alpha_{yy}$ and
$\sigma_{xx}\alpha_{xy}\gg\sigma_{xy}\alpha_{xx}$.
As a result, we have $S_{xy}/S_{yx}\approx\sigma_{yy}/\sigma_{xx}$
from Eqs. (8) and (9).
Due to the obvious anisotropy of the longitudinal conductivity
$\sigma_{xx}<\sigma_{yy}$, we can conclude that the Nernst signal
is also anisotropic, $S_{xy}>S_{yx}$.

\begin{figure}[tbh]
\includegraphics[width=2.9in]{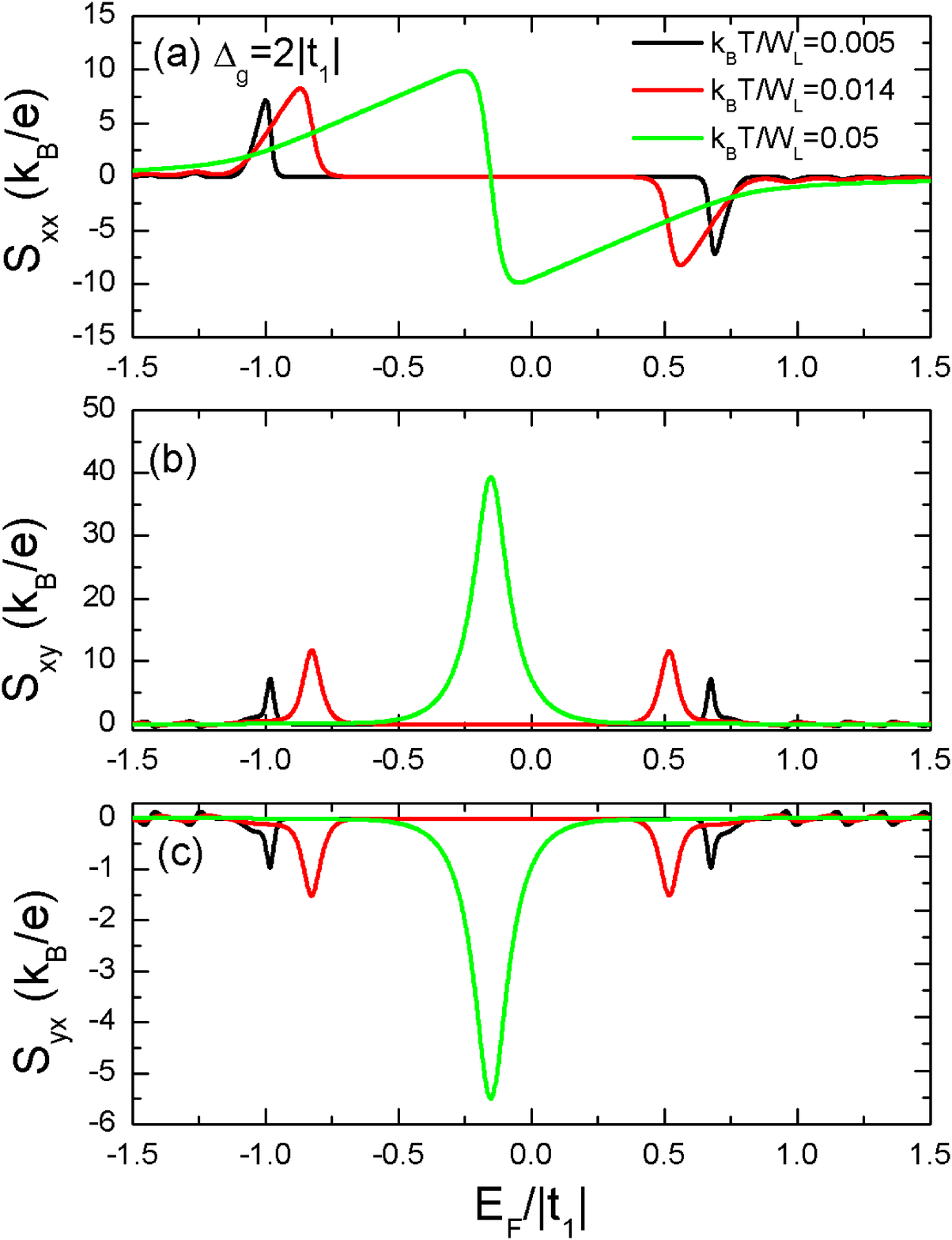}
\caption{ (color online). Calculated thermopower and Nernst signal
%as functions of the Fermi energy
in the presence of an applied bias voltage $\Delta_g$ of phosphorene at three different temperatures.
} \label{fig.6}
\end{figure}

More interesting, when a bias voltage $\Delta_g$ is applied between
the top and bottom layers of phosphorene, the peak values of thermopower
and Nernst signal become large.
As seen from Figs.\ref{fig.6}(a), the peak values of thermopower $S_{xx}$
increase to $\pm 9.9 k_B/e$ ($\pm 853.1 \mu V/K$),
which are greater than those in the unbiased case.
The enhanced thermopower is mainly due to the increase of
the bulk energy gap between the valence and conduction bands.
According to the definition of the thermopower, $S_{xx}$
is determined by the electron-transmission-weighted average
value of the heat energy $E$-$E_F$.
Due to the increase of the bulk energy gap in biased phosphorene, the electrons
near the conduction band edge, which are responsible for the maximum thermopower,
have a much larger $E$-$E_F$ compared to the case of unbiased phosphorene.
This is similar to the situation in semiconducting
armchair graphene nanoribbons~\cite{ouyang2009,xing2009,hao2010}.
On the other hand,
the peak value of $S_{xy}$ increases to $39.4 k_B/e$
(3395.1 $\mu V/K$), and the value of $S_{yx}$ increases to
$-5.5$ $k_B/e$ ($-473.9$ $\mu V/K$).
The enhanced thermopower and Nernst signal are very beneficial
for the thermoelectric applications of phosphorene-based materials.
It is known that a large thermopower does not necessarily lead to an
enhanced power factor.
On the contrary, a moderate thermopower combined with a suitable electrical conductivity
may eventually result in a high power factor.
The similar phenomenon has also been observed in experiments~\cite{Zhao2014}.

\begin{figure}[tbh]
\includegraphics[width=2.9in]{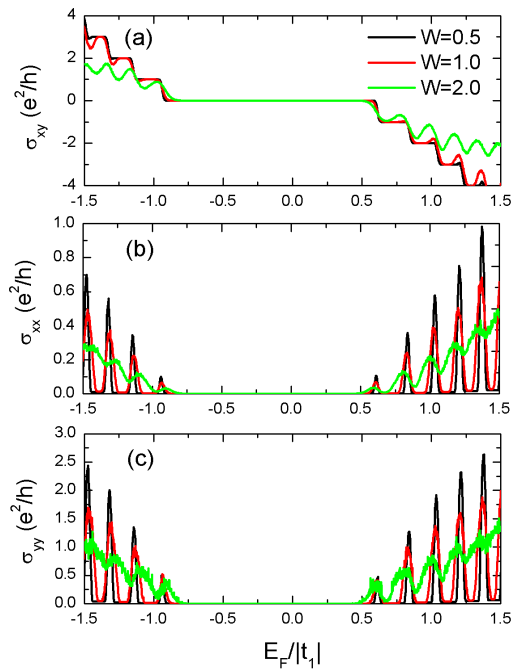}
\caption{ (color online). Calculated Hall conductivity $\sigma_{xy}$ and
longitudinal conductivity $\sigma_{xx}$ ($\sigma_{yy}$) in units of $e^2/h$
as functions of the Fermi energy at zero temperature in phosphorene
for three different disorder strengths.
%The other parameters are chosen to be the same as in Fig.\ref{fig.2}.
} \label{fig.7}
\end{figure}

\subsection{Disorder effect on the electrical and thermoelectric transport}

Now we study the effect of disorder on the electrical conductivity
in phosphorene. In Fig.\ref{fig.7}, both Hall conductivity $\sigma _{xy}$
and longitudinal conductivity $\sigma _{xx}$ are shown as functions of $E_F$
for three different disorder strengths.
As seen from Fig.\ref{fig.7}(a), the plateaus with
$\nu =0, \pm 1,\pm 2$ and $\pm 3$ remain well quantized at $W=0.5$.
With increasing $W$, the higher Hall plateaus (with larger $|\nu |$)
are destroyed first because of the relatively small plateau widths.
At $W=2.0$, only the $\nu =0$ QHE state remains robust.
Clearly, after the destruction of the QHE states near the band edge,
all the electron states become localized.
Then the topological Chern numbers initially carried by these states will
move towards band center in a similar manner to the case of graphene~\cite{Sheng2006}.
Thus the phase diagram indicates a float-up picture, in which the extended
levels move towards band center with increasing disorder strength,
causing higher plateaus to disappear first.
As seen from Figs.\ref{fig.7}(b)and (c), the peaks of the calculated
longitudinal conductivity $\sigma _{xx}$ are strongly broadened
with increasing disorder strength.

\begin{figure}[tbh]
\includegraphics[width=2.6in]{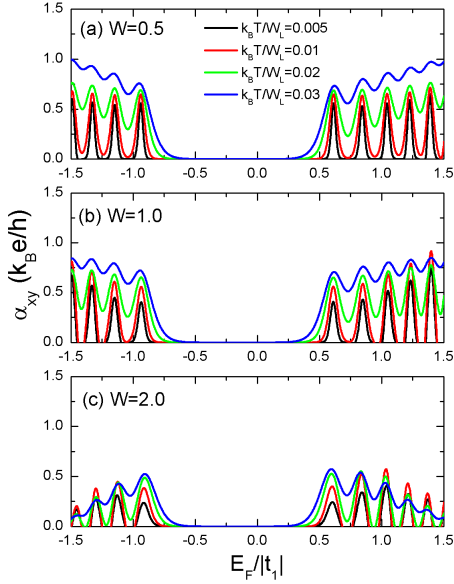}
\caption{ (color online). The transverse thermoelectric
conductivity $\alpha_{xy}$ of phosphorene for three different disorder strengths.
(a)$W=0.5$, (b)$W=1.0$ and (c)$W=2.0$. Here, the asymmetric gaps $W_L$
are equal to $W_L/\vert t_1 \vert = 1.851, 1.824,$ and $1.755$, respectively.
} \label{fig.8}
\end{figure}

\begin{figure}[tbh]
\includegraphics[width=3.7in]{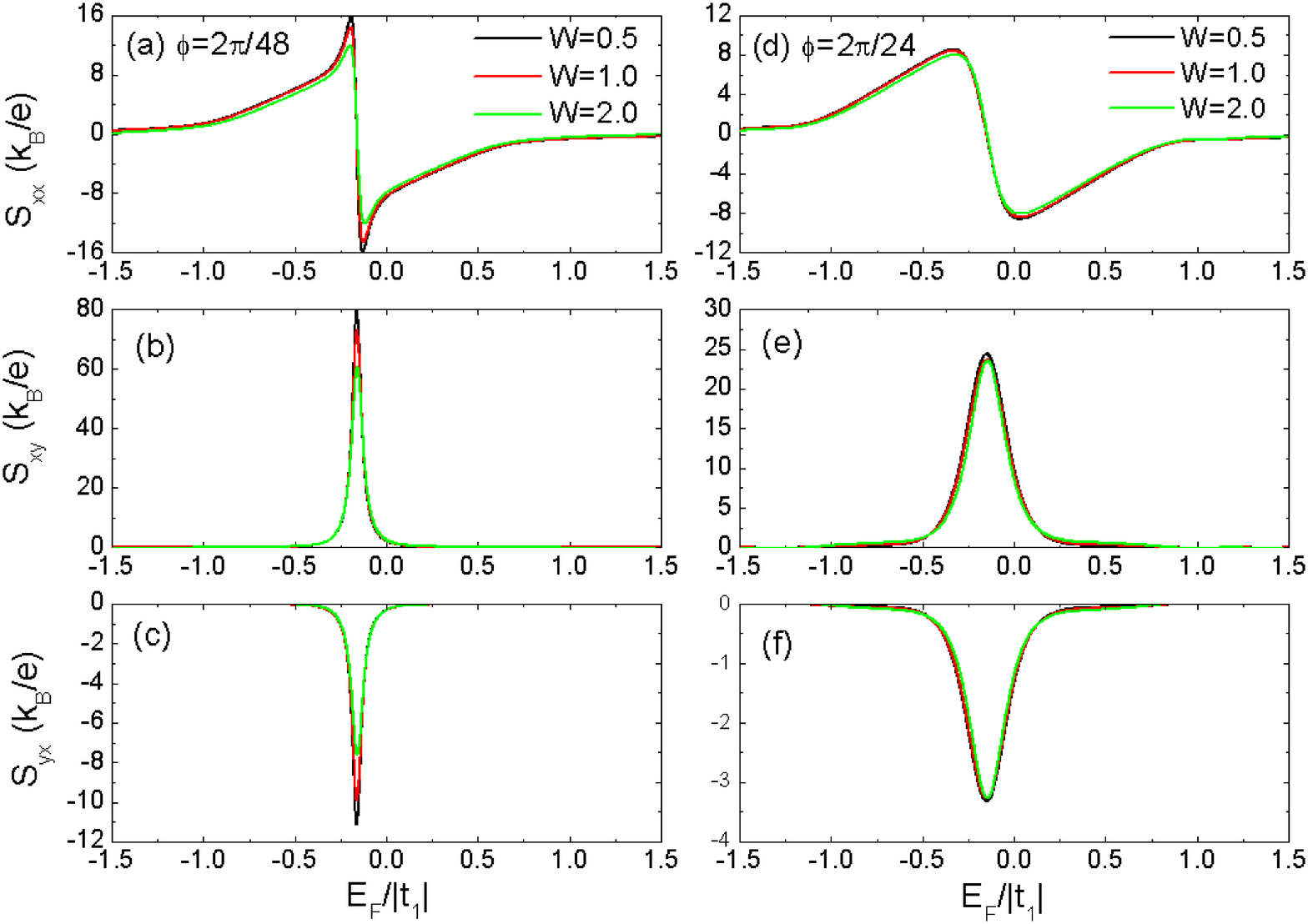}
\caption{ (color online). Calculated thermopower and Nernst signal of phosphorene
for three different disorder strength at a fix temperature $k_BT=0.05W_L$.
Here, $W_L$ is chosen as $W_L/\vert t_1 \vert = 1.8$.
(a)-(b)$\phi=2\pi/48$, (c)-(d)$\phi=2\pi/24$.
%The other parameters are chosen to be the same as Fig.\ref{fig.7}.
} \label{fig.9}
\end{figure}

We also investigate the disorder effect on the thermoelectric conductivity
in phosphorene. In Fig.\ref{fig.8}, the transverse thermoelectric
conductivity $\alpha_{xy}$ for three different disorder strengths
$W=$0.5, 1.0 and 2.0 is shown. It is found that $\alpha_{xy}$ displays
a series of peaks at the center of each LL.
With increasing disorder strength from $W=0.5$ to $W=2.0$,
the widths of peaks in $\alpha_{xy}$ increase.
All the peaks will disappear around $W\sim 3.0$, which is caused by
the merging of states with opposite Chern numbers at
strong disorder~\cite{Sheng2006}.

We finally investigate the disorder effect on the thermopower
and Nernst signal in phosphorene.
In Figs. \ref{fig.9}(a)-(c), the calculated $S_{xx}$, $S_{xy}$ and $S_{yx}$
are plotted for different disorder strengths with magnetic flux $\phi=2\pi/48$.
It is well known that when the magnetic field is absent,
the thermopower is strongly affected by the disorder, and the peaks
are suppressed even for small disorder~\cite{Ma2016}.
However, in the presence of the strong magnetic field, both thermopower
and Nernst signal are robust to the disorder, due to the fact that
the highly-degenerated LLs dominate transport processes.
When the magnetic field is increased to $\phi=2\pi/24$,
it is interesting to find that the peak heights of $S_{xx}$, $S_{xy}$ and $S_{yx}$
remain almost unchanged with increasing the disorder strength,
as seen from Figs. \ref{fig.9}(d)-(f).
It means that, the stronger the magnetic field is, the more robust
the thermopower and Nernst signal.
In fact, a similar conclusion has been reached in the study of
disorder effect in graphene nanoribbons~\cite{xing2009}.

\section{Summary}

In summary, we have numerically investigated the electrical and
thermoelectric transport properties of phosphorene in the presence of
both a magnetic field and disorder.
The quantized Hall conductivity is similar to that of a conventional
2DEG, but the positions of all the Hall plateaus shift to the left
due to the spectral asymmetry.
The thermoelectric conductivities and Nernst signal
exhibit remarkable anisotropy, and the thermopower is nearly isotropic.
Upon applying a bias voltage to phosphorene,
the quantized Hall plateaus remain to follow the same sequence,
but the width of $\nu=0$ plateau increases.
It is interesting to find that the peak values of the thermopower and
Nernst signal become larger compared to the unbiased case.
We attribute the large magnitude of the thermopower to the increase of
the bulk energy gap. Moreover, we also study the disorder effect
on the electrical and thermoelectric transport in phosphorene.
With increasing disorder strength, the Hall plateaus can be destroyed
through the float-up of extended levels toward the band center and
higher plateaus disappear first. The $\nu=0$ plateau is most robust against
disorder scattering. In the presence of the strong magnetic field,
both thermopower and Nernst signal are robust to the disorder,
because of the existence of the quantized LLs.
The stronger the magnetic field is,
the more robust the thermopower and Nernst signal.

\acknowledgments This work was supported by the National Natural Science Foundation
of China under grant numbers 11574155, 11681240385 (R.M.), and
11674160 (L.S.). This work was also supported
by the State Key Program for Basic Researches of China under grant numbers
2015CB921202, 2014CB921103 (L.S.) and a project funded by China Postdoctoral
Science Foundation under grant numbers 2014M551546, 2015T80532(R.M.).
This work was also supported by the U.S. Department of Energy,
Office of Basic Energy Sciences under grants No. DE-FG02-
06ER46305(D. N. Sheng).


\begin{references}
\bibitem{footnote1} Different literatures may have a sign difference due to different conventions.
\end{references}
\end{document}